# Inducing Quantum Phase Transitions in Non-Topological Insulators Via Atomic Control of Sub-Structural Elements

Thomas K. Reid, [1] S. Pamir Alpay, [1,2] Alexander V. Balatsky, [2,3] and Sanjeev K. Nayak [1,2,*]

[1] Department of Materials Science and Engineering and Institute of Materials Science, University of Connecticut, Storrs, Connecticut 06269, USA

[2] Department of Physics, University of Connecticut, Storrs, CT 06269, USA

[3] NORDITA, KTH Royal Institute of Technology and Stockholm University, Stockholm SE-106 91, Sweden

## ABSTRACT

Topological insulators (TIs) are an important family of quantum materials that exhibit a Dirac point (DP) in the surface band structure but have a finite band gap in bulk. A large degree of spin-orbit interaction and low bandgap is a prerequisite for stabilizing DPs on selective atomically flat cleavage planes. Tuning of the DP in these materials has been suggested via modifications to the atomic structure of the entire system. Using the example of $As_2Te_3$ and $ZnTe_5$, which are not TIs, we show that a quantum phase transition can be induced in atomically flat and stepped surfaces, for $As_2Te_3$ and $ZrTe_5$, respectively. This is achieved by establishing a framework for controlling electronic properties that is focused on local perturbations at key locations that we call sub-structural elements (SSEs). We exemplify this framework through a novel method of isovalent sublayer anion doping and biaxial strain.

[*] Corresponding author: sanjeev.nayak@uconn.edu

Topological insulators (TIs) are materials that have topologically protected surface states characterized by the Dirac point (DP) that is robust to non-time reversal perturbations [1–3]. A large degree of spin-orbit coupling (SOC) in these materials leads to entanglement of band extrema at the Fermi level, leading to interchange of atomic contributions to the band occupancy in the bulk, as compared to regular semiconductors. This effect manifests on the surface as a DP, which is characterized by a zero bandgap, linear band dispersion, and momentum-spin locking of chiral spin order. Graphene is the first material where a DP was observed, however for active applications, functionalized graphene is in under active consideration [4,5]. TIs have the advantage of a richer chemical space to engineer their properties. Projected applications of TIs include superconducting materials, quantum computers, spintronic devices, and quantum anomalous Hall insulators, making it a fruitful topic for investigation [6–8].

Realizing these quantum effects in naturally existing materials is a major challenge. Thus, much of TI research is focused on either conceptualizing new materials or tuning the properties of known materials [9]. Initial studies on the physics of TIs centered around the binary AB-type or $A_2B_3$-type chalcogenides, which laid fertile groundwork for studies that continue on this class of materials today [10–21]. Density Functional Theory (DFT) modeling with the scalar-relativistic Kohn-Sham Hamiltonian and perturbative SOC is sufficient to account for the surface physics of target materials [5,22]. As a consequence, many of these modern studies depend upon the use of DFT-based computational methods.

Recently, we have applied various density-functional approximations for the exchange-correlation term to a representative sample of the $A_2B_3$ binary pnictogenide chalcogenide (BPC) TI group. Because of the quintuple layer (QL) structure of these materials, it is also necessary to consider the role of the van der Waals (vdW) interaction, which we incorporated via the dispersion-corrected parameters [23]. We found a set of methodological rules that are generally suitable for the $A_2B_3$ group. These are: the use of a slab model with six or more unit-layers, the use of a vacuum thickness $\geq$ 15 Å, the use of the generalized gradient approximation with the vdW parameterization (GGA + vdW), inclusion of SOC, and thorough optimization of geometry.

Using the above technique, we have shown that thermoelectric non-TI semiconductor $As_2Te_3$ undergoes a quantum phase transition, becoming a TI, when subject to 1% strain in the *ab*-plane. This is due to a Poisson effect, whereby the slab thickness undergoes a reduction in response to the extension of *a* and *b*; we further demonstrated the disproportionate contribution of the vdW gaps ($d_{vdW}$), which are "stacked" along *c*, in producing this reduction [24,25]. In addition, a correlate can be established with reduction in the outer atomic layer separation ($d_{OAL}$) toward control of TI functionality. Our study has shown that $d_{OAL}$ can vary up to 11% for under conditions of ±3% strain, which matches well to experiments performed on surface C-doped $Bi_2Se_3$, which report 10.7% variation [26,27]. Control over the TI properties was also observed in the $Pb_{1-x}Sn_xTe$ system using the GGA exchange-correlation functional, which possesses NaCl structure throughout the composition range $0 \leq x \leq 1$, via application of hydrostatic strain [28].

Given that biaxial strain can only be readily induced by substrate lattice-mismatch, the degree of strain is naturally dependent on the selection of a suitably mismatched substrate. A meaningful evolution of the standard method of analyzing the atomic structure of BPCs, which has primarily focused on "whole-system" structural features such as the number of quintuple layers (QLs), the overall thickness, and the orientation of the sample can be found by focusing upon parameters such as the outermost $d_{vdW}$ ($d_{OvdW}$) and $d_{OAL}$, which has important implications. In the $As_2Te_3$ case, by focusing on $d_{OvdW}$ alone, we were able to observe that a single sub-feature of the atomic structure could have an outsize, and indeed primary, impact on the bandgap [25,29]. It thus became apparent that bandgap tuning in BPCs could be made possible by careful mediation of sub-structural elements (SSEs) like the $d_{OvdW}$ and $d_{OAL}$. While external strain factors into wholesome mechanical response to the electronic structure, an internal and localized "chemical" strain is an equally powerful method to modifying materials properties, and is not bound by such restrictions [28]. In this aspect, doping is a viable option and, with respect to the given problem, one looks forward to controlling the interlayer separation by isovalent doping to minimize changes to the bonding type. Overall, its advantages are numerous. Firstly, it introduces structural changes intrinsically, avoiding dependencies like lattice-mismatch strain. Secondly, it provides control over the chemistry of the system. And lastly, the structural perturbations are of local order.

Subsequently, we sought to explore this premise in two ways: a) finding a method that demonstrates the feasibility of modifying SSEs in practice and b) demonstrating the primacy of certain SSEs in determining electronic properties, like the bandgap, in layered TI and TI-adjacent (DFT) together with spin-orbit interaction and report success in both pursuits [24]. Altogether, we show that a novel approach of sublayer isovalent anion doping (SIAD) and controlling dopant concentration offers major control over properties-relevant SSEs in $As_2Te_3$, and that the sublayer doping approach could be extended to stepped surfaces. For the stepped surface case, we chose to work on ZrTe$_5$, in which we achieved total bandgap closure by substitutional doping at singular atomic sites.

In previous work, we studied the 6 QL As$_2$Te$_3$ slab constructed from the $R\bar{3}m$ unit-cell [25]. Now, we present the results of the application of our doping strategy to this material. First, we built a set of models corresponding to sixteen doping scenarios. For the chemistry of the dopant, we designated tellurium-isovalent elements S and Se, which we substituted at the fifth atomic layer (the lowest atomic layer of the outermost QL) and the sixth atomic layer (the highest atomic layer of the second QL). This is schematically illustrated in Figure 1(a). Each As$_2$Te$_3$ slab possesses a 2×2 reciprocity in the *ab*-plane, which served to assess the effect of dopant concentration. For both dopants, at both the fifth and sixth layer, tellurium substitution proceeded progressively, with a model created for one, two, three, and four tellurium atoms replaced.

We examined a set of SSEs in comparison to the dopant site and concentration: d$_{OvdW}$, the thickness of the outermost QL (d$_{OQL}$), and the separation between the two outermost atomic layers (d$_{OAL}$). In order to simplify the comparison between the original and doped structures, we refer to the change of d$_{OvdW}$ and d$_{OAL}$ (respectively, Δd$_{OvdW}$ and Δd$_{OAL}$). This is shown in Table 1. What is apparent is a major reduction in all three SSEs in comparison to the undoped As$_2$Te$_3$ slab, with the Δd$_{OvdW}$ showing a particularly large negative response to sulfur in layer five at concentrations exceeding 75%. This would appear to demonstrate the utility of substitutional anion doping for controlling BPC atomic structure. A major reduction in the bandgap from the undoped structure from 0.130 eV to −0.035 eV is achieved when sulfur is at 100% concentration in layer 5, with the other doping conditions inducing reductions of the bandgap of various degrees that are less severe. The overall picture is clear: by targeting the anions surrounding d$_{OvdW}$, significant control over the vdW gap and surrounding SSEs can be exerted, which in turn exerts major negative pressure on the bandgap, up to and including total bandgap closure. Given that As$_2$Te$_3$ is a non-TI, we have further demonstrated that doping presents a plausible method for the transformation of non-TI BPCs into TIs, facilitated by an understanding of properties-relevant SSEs.

Due to the challenges currently surrounding its characterization as a TI, we decided to extend the hypothesis to ZrTe$_5$, which also possesses a distinct structure and stoichiometry. To accomplish this, we first constructed a set of thirty-six ZrTe$_5$ slab models, corresponding to eighteen tellurium doping layer-sites, for both sulfur and selenium dopants, each with 12-atom ZrTe$_5$ prisms. Each model, therefore, reflects substitutional doping of tellurium at one of the sites, with either sulfur or selenium. These doping layer-sites proceed from the "top" of the material to the tellurium layer just "above" the central van der Waals gap (parallel to *b*-axis). All models possess 2 × 1 *ac*-plane reciprocation, and, at each layer-site, 25% of the tellurium anions were replaced. These doping scenarios are schematically illustrated in Fig. 1 (b). It must be noted that the ZrTe$_5$ (010) surface is "stepped", while the BPC (0001) surface is flat.

In Fig. 2 (a), we show the relationship between the bandgap of doped ZrTe$_5$, the layer-site doped, and the dopant used. Firstly, we note that pristine, six-layer ZrTe$_5$ is a non-TI that maintains a bandgap of 0.32 meV. This has been experimentally validated although the bandgap value reported was lower than that we predicted [30,31]. It is apparent that a major negative response in the bandgap occurs—total bandgap closure, in nearly every case—when a sulfur dopant is located at layers $(3 + 6n)$ or $(4 + 6n) \forall n \in 0, 1,$ and 2—these positions in the structure correspond to the sites in the ZrTe$_5$ structure which we designate the central anion pairs (CAP), which are highlighted in Figure 1 (b). Conversely, we observe significant bandgap opening with selenium doping. As such, we identify the CAP sites as a high-importance SSE in ZrTe$_5$ for tuning electronic properties. Figure 2 (b) emphasizes this point, showing a strong relationship between the layer-site doped, and post-relaxation changes to the various SSEs and the total thickness of the slab. Notably, the nearest-neighbor S–Zr bond length along the layers is smaller than nearest-neighbor Se–Zr bond length. However, the CAP sites, where the *X*–Zr bond lengths are larger than other layers, are opposite in trend as compared to the undoped (initial) structure. By replacing a single anion in a structure of 144 atoms, we are able to achieve significant control of the bandgap, so long as we target the correct SSE [see Figure 3 (b)]. We take special note that, in the case of Te-layer-site $(3 + 6n)$ and $(4 + 6n)$ sulfur doping, total or near-total bandgap closure, a clear Dirac-like dispersion near the Gamma point, and occupancy inversion is achieved according to the criteria of our previous work (see Refs. [24,28]).

In summary, in contrast to a focus on whole-system properties like the total thickness and global strain, the use of high-impact SSEs facilitates materials design through targeted techniques like SIAD. In this study, the general importance of identification and control of SSEs for both As$_2$Te$_3$ and ZrTe$_5$ is clearly supported. Through the application of our doping strategy, we are able to achieve major control over the bandgap of both materials with either atomic layer or single-atom substitutions at key structural sites. Specifically, we identified the d$_{OvdW}$ and CAP sites for As$_2$Te$_3$ and ZrTe$_5$, respectively (see Figure 1). In the ZrTe$_5$ case, this matches well to previous experimental and theoretical studies. It should be noted that, in both materials, it is sulfur that achieves bandgap closure. In general, we see a sensitivity of the topological surface state to various conditions like strain and temperature [30–35]. Systematic DFT analysis of the SSEs for a system helps to identify ones that are important for electronic control.

Our study emphasizes the potential for SSEs in TI and TI-adjacent materials to serve as a vehicle for the tuning of electronic properties. This can be achieved experimentally with advanced synthesis techniques like molecular beam epitaxy, selective ion implantation, and scanning-tunneling microscopy. Such an approach can be complimented by methods for modifying whole-system properties like biaxial strain to achieve an exact-zero bandgap. In addition, due to the novelty and experimental accessibility of the SSE-SIAD framework, we also verify the utility of a method for control over the electronic state of TIs and TI-adjacent materials, with hitherto unexplored potential in an experimental setting. Together, these insights indicate a promising new course for the study and development of next generation of TI-based applications.

**DATA AVAILABILITY**

The data that supports the findings of this study are available within the article.

**ACKNOWLEDGEMENTS**

The authors are grateful for the computational resources provided by the University of Connecticut. This work was supported by European Research Council under the European Union Seventh Framework ERS-2018-SYG 810451 HERO, the Knut and Alice, Wallenberg Foundation KAW 2019.0068, and the University of Connecticut.

# REFERENCES

[1] B. Bradlyn, L. Elcoro, J. Cano, M. G. Vergniory, Z. Wang, C. Felser, M. I. Aroyo, and B. A. Bernevig, *Topological Quantum Chemistry*, Nature **547**, 298 (2017).

[2] M. Z. Hasan and C. L. Kane, Colloquium *: Topological Insulators*, Rev. Mod. Phys. **82**, 3045 (2010).

[3] S. Gupta and A. Saxena, *A Topological Twist on Materials Science*, MRS Bull. **39**, 265 (2014).

[4] S. Sahoo, S. L. Suib, and S. P. Alpay, *Graphene Supported Single Atom Transition Metal Catalysts for Methane Activation*, ChemCatChem **10**, 3229 (2018).

[5] S. Sahoo, M. E. Gruner, S. N. Khanna, and P. Entel, *First-Principles Studies on Graphene-Supported Transition Metal Clusters*, J. Chem. Phys. **141**, 74707 (2014).

[6] T. O. Wehling, A. M. Black-Schaffer, and A. V Balatsky, *Dirac Materials*, Adv. Phys. **63**, 1 (2014).

[7] X.-L. Qi and S.-C. Zhang, *Topological Insulators and Superconductors*, Rev. Mod. Phys. **83**, 1057 (2011).

[8] F. Duncan and M. Haldane, *Topological Quantum Matter*, Int. J. Mod. Phys. B **32**, (2018).

[9] G. Aeppli, A. V Balatsky, H. M. Rønnow, and N. A. Spaldin, *Hidden, Entangled and Resonating Order*, Nat. Rev. Mater. **5**, 477 (2020).

[10] X. Yang, L. Luo, C. Vaswani, X. Zhao, Y. Yao, D. Cheng, Z. Liu, R. H. J. Kim, X. Liu, M. Dobrowolska-Furdyna, J. K. Furdyna, I. E. Perakis, C. Wang, K. Ho, and J. Wang, *Light Control of Surface–Bulk Coupling by Terahertz Vibrational Coherence in a Topological Insulator*, Npj Quantum Mater. **5**, 13 (2020).

[11] D. Flötotto, Y. Bai, Y.-H. Chan, P. Chen, X. Wang, P. Rossi, C.-Z. Xu, C. Zhang, J. A. Hlevyack, J. D. Denlinger, H. Hong, M.-Y. Chou, E. J. Mittemeijer, J. N. Eckstein, and T.-C. Chiang, *In Situ Strain Tuning of the Dirac Surface States in $Bi_2Se_3$ Films*, Nano Lett. **18**, 5628 (2018).

[12] M. Safdar, Q. Wang, M. Mirza, Z. Wang, K. Xu, and J. He, *Topological Surface Transport Properties of Single-Crystalline Snte Nanowire*, Nano Lett. **13**, 5344 (2013).

[13] T. H. Hsieh, H. Lin, J. Liu, W. Duan, A. Bansi, and L. Fu, *Topological Crystalline Insulators in the SnTe Material Class*, Nat. Commun. **3**, 982 (2012).

[14] S. M. Young, S. Chowdhury, E. J. Walter, E. J. Mele, C. L. Kane, and A. M. Rappe, *Theoretical Investigation of the Evolution of the Topological Phase of Bi2Se3 under Mechanical Strain*, Phys. Rev. B **84**, 085106 (2011).

[15] S. S. Hong, W. Kundhikanjana, J. J. Cha, K. Lai, D. Kong, S. Meister, M. A. Kelly, Z.-X. Shen, and Y. Cui, *Ultrathin Topological Insulator $Bi_2Se_3$ Nanoribbons Exfoliated by Atomic Force Microscopy*, Nano Lett. **10**, 3118 (2010).

[16] Y. Zhang, K. He, C.-Z. Chang, C.-L. Song, L.-L. Wang, X. Chen, J.-F. Jia, Z. Fang, X. Dai, W.-Y. Shan, S.-Q. Shen, Q. Niu, X.-L. Qi, S.-C. Zhang, X.-C. Ma, and Q.-K. Xue, *Crossover of the Three-Dimensional Topological Insulator $Bi_2Se_3$ to the Two-Dimensional Limit*, Nat. Phys. **6**, 584 (2010).

[17] P. Tsipas, E. Xenogiannopoulou, S. Kassavetis, D. Tsoutsou, E. Golias, C. Bazioti, G. P.

Dimitrakopulos, P. Komninou, H. Liang, M. Caymax, and A. Dimoulas, *Observation of Surface Dirac Cone in High-Quality Ultrathin Epitaxial $Bi_2Se_3$ Topological Insulator on AlN(0001) Dielectric*, ACS Nano **8**, 6614 (2014).

[18] H. Zhang, C.-X. Liu, X.-L. Qi, X. Dai, Z. Fang, and S.-C. Zhang, *Topological Insulators in $Bi_2Se_3$, $Bi_2Te_3$ and $Sb_2Te_3$ with a Single Dirac Cone on the Surface*, Nat. Phys. **5**, 438 (2009).

[19] C. Schindler, C. Wiegand, J. Sichau, L. Tiemann, K. Nielsch, R. Zierold, and R. H. Blick, *Strain-Induced Dirac State Shift in Topological Insulator $Bi_2Se_3$ Nanowires*, Appl. Phys. Lett. **111**, 171601 (2017).

[20] J. Liu, W. Duan, and L. Fu, *Two Types of Surface States in Topological Crystalline Insulators*, Phys. Rev. B **88**, 241303(R) (2013).

[21] Y. Tanaka, Z. Ren, T. Sato, K. Nakayama, S. Souma, T. Takahashi, K. Segawa, and Y. Ando, *Experimental Realization of a Topological Crystalline Insulator in SnTe*, Nat. Phys. **8**, 800 (2012).

[22] S. Sahoo, A. Hucht, M. E. Gruner, G. Rollmann, P. Entel, A. Postnikov, J. Ferrer, L. Fernández-Seivane, M. Richter, D. Fritsch, and S. Sil, *Magnetic Properties of Small Pt-Capped Fe, Co, and Ni Clusters: A Density Functional Theory Study*, Phys. Rev. B **82**, 54418 (2010).

[23] S. Grimme, J. Antony, S. Ehrlich, and H. Krieg, *A Consistent and Accurate* Ab Initio *Parametrization of Density Functional Dispersion Correction (DFT-D) for the 94 Elements H-Pu*, J. Chem. Phys. **132**, 154104 (2010).

[24] T. K. Reid, S. P. Alpay, A. V. Balatsky, and S. K. Nayak, *First-Principles Modeling of Binary Layered Topological Insulators: Structural Optimization and Exchange-Correlation Functionals*, Phys. Rev. B **101**, 085140 (2020).

[25] T. K. Reid, S. K. Nayak, and S. P. Alpay, *Strain-Induced Surface Modalities in Pnictogen Chalcogenide Topological Insulators*, J. Appl. Phys. **129**, 15304 (2021).

[26] S. Roy, H. L. Meyerheim, A. Ernst, K. Mohseni, C. Tusche, M. G. Vergniory, T. V Menshchikova, M. M. Otrokov, A. G. Ryabishchenkova, Z. S. Aliev, M. B. Babanly, K. A. Kokh, O. E. Tereshchenko, E. V Chulkov, J. Schneider, and J. Kirschner, *Tuning the Dirac Point Position in $Bi_2Se_3$(0001) via Surface Carbon Doping*, Phys. Rev. Lett. **113**, 116802 (2014).

[27] S. Roy, H. L. Meyerheim, K. Mohseni, A. Ernst, M. M. Otrokov, M. G. Vergniory, G. Mussler, J. Kampmeier, D. Grützmacher, C. Tusche, J. Schneider, E. V Chulkov, and J. Kirschner, *Atomic Relaxations at the (0001) Surface of $Bi_2Se_3$ Single Crystals and Ultrathin Films*, Phys. Rev. B **90**, 155456 (2014).

[28] M. Geilhufe, S. K. Nayak, S. Thomas, M. Däne, G. S. Tripathi, P. Entel, W. Hergert, and A. Ernst, *Effect of Hydrostatic Pressure and Uniaxial Strain on the Electronic Structure of $Pb_{1-x}Sn_xTe$*, Phys. Rev. B **92**, 235203 (2015).

[29] K. Shirali, S. W A, and I. Vekhter, *Inter-Quintuple Layer Coupling and Topological Phase Transitions in the Chalcogenide Topological Insulators*, Electron. Struct. **5**, 15001 (2023).

[30] I. Mohelsky, J. Wyzula, B. A. Piot, G. D. Gu, Q. Li, A. Akrap, and M. Orlita, *Temperature Dependence of the Energy Band Gap in $ZrTe_5$: Implications for the Topological Phase*, Phys. Rev. B **107**, L041202 (2023).

[31] J. Mutch, W.-C. Chen, P. Went, T. Qian, I. Z. Wilson, A. Andreev, C.-C. Chen, and J.-H. Chu, *Evidence for a Strain-Tuned Topological Phase Transition in ZrTe$_5$*, Sci. Adv. **5**, (2019).

[32] J. Wang, Y. Jiang, T. Zhao, Z. Dun, A. L. Miettinen, X. Wu, M. Mourigal, H. Zhou, W. Pan, D. Smirnov, and Z. Jiang, *Magneto-Transport Evidence for Strong Topological Insulator Phase in ZrTe$_5$*, Nat. Commun. **12**, 6758 (2021).

[33] H. Xiong, J. A. Sobota, S.-L. Yang, H. Soifer, A. Gauthier, M.-H. Lu, Y.-Y. Lv, S.-H. Yao, D. Lu, M. Hashimoto, P. S. Kirchmann, Y.-F. Chen, and Z.-X. Shen, *Three-Dimensional Nature of the Band Structure of ZrTe$_5$ Measured by High-Momentum-Resolution Photoemission Spectroscopy*, Phys. Rev. B **95**, 195119 (2017).

[34] N. Aryal, X. Jin, Q. Li, A. M. Tsvelik, and W. Yin, *Topological Phase Transition and Phonon-Space Dirac Topology Surfaces in ZnTe$_5$*, Phys. Rev. Lett. **126**, 16401 (2021).

[35] P. Zhang, R. Noguchi, K. Kuroda, C. Lin, K. Kawaguchi, K. Yaji, A. Harasawa, M. Lippmaa, S. Nie, H. Weng, V. Kandyba, A. Giampietri, A. Barinov, Q. Li, G. D. Gu, S. Shin, and T. Kondo, *Observation and Control of the Weak Topological Insulator State in ZrTe$_5$*, Nat. Commun. **12**, 406 (2021).

**Table 1**. The bandgap compared to structural parameters for S-doped $As_2Te_3$ models for different concentration of S in Te-layer-sites 5 and 6 [see Fig. 1 (a)]. The structural parameters are presented in Δ, which is computed as geometrically optimized state with dopants minus optimized state of the corresponding pure model.

| Layer concentration (%) | Bandgap (eV) | $\Delta d_{OQL}$ (Å) | $\Delta d_{OAL}$ (Å) | $\Delta d_{OvdW}$ (Å) |
|---|---|---|---|---|
| **Doping $S_{Te}$ on layer 5** | | | | |
| 0 | 0.130 | – | – | – |
| 0.25 | 0.105 | 0.019 | 0.011 | −0.212 |
| 0.50 | 0.059 | 0.144 | −0.134 | −0.365 |
| 0.75 | 0.060 | 0.207 | 0.039 | −0.467 |
| 1.00 | −0.032 | −0.432 | 0.032 | −0.431 |
| **Doping $S_{Te}$ on layer 6** | | | | |
| 0 | 0.130 | – | – | – |
| 0.25 | 0.118 | −0.005 | 0.103 | −0.188 |
| 0.50 | 0.109 | −0.003 | 0.070 | −0.354 |
| 0.75 | 0.129 | −0.003 | 0.069 | −0.456 |
| 1.00 | 0.012 | −0.011 | 0.035 | −0.363 |
| **Doping $Se_{Te}$ on layer 5** | | | | |
| 0 | 0.130 | – | – | – |
| 0.25 | 0.107 | 0.006 | 0.017 | −0.141 |
| 0.50 | 0.102 | 0.068 | −0.074 | −0.296 |
| 0.75 | 0.084 | 0.111 | 0.136 | −0.401 |
| 1.00 | 0.032 | −0.247 | 0.033 | −0.351 |
| **Doping $Se_{Te}$ on layer 6** | | | | |
| 0 | 0.130 | – | – | – |
| 0.25 | 0.115 | −0.011 | 0.081 | −0.138 |
| 0.50 | 0.118 | −0.005 | 0.064 | −0.280 |
| 0.75 | 0.127 | 0.014 | 0.064 | −0.414 |
| 1.00 | 0.090 | −0.013 | 0.032 | −0.324 |

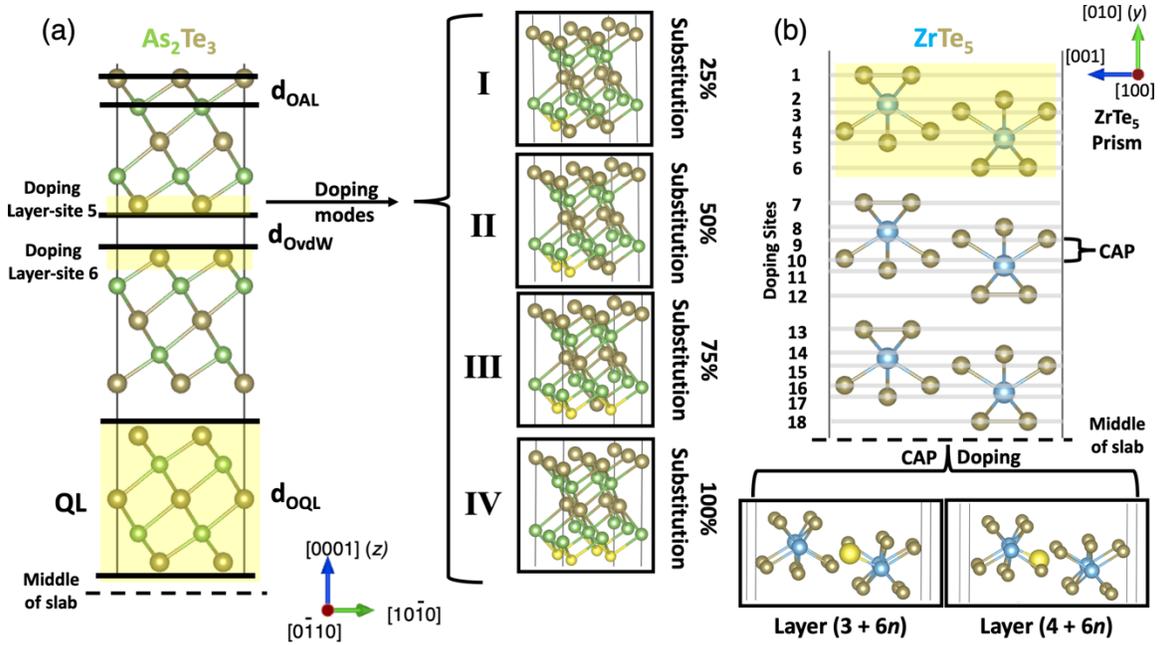

**Figure 1.** Schematic representations of the As$_2$Te$_3$ (a) and ZrTe$_5$ (b) slab models are shown. Various structural parameters, defined according to our SSE-SIAD framework, are illustrated in relation to the structure, along with other features of note. In (a), the "Doping Layer-site" labels correspond to the layers of Te atoms that were substituted to explore the effect of SIAD on the band structure, which are highlighted in yellow for additional clarity and to distinguish them from the SSEs $d_{OAL}$, $d_{OvdW}$, and $d_{OQL}$. In (b), CAP is identified at "Doping Sites" 9 and 10, but is present within all three ZrTe$_5$ "Prisms" studied at layers $(3 + 6n)$ or $(4 + 6n)$ $\forall\, n \in 0, 1,$ and 2.

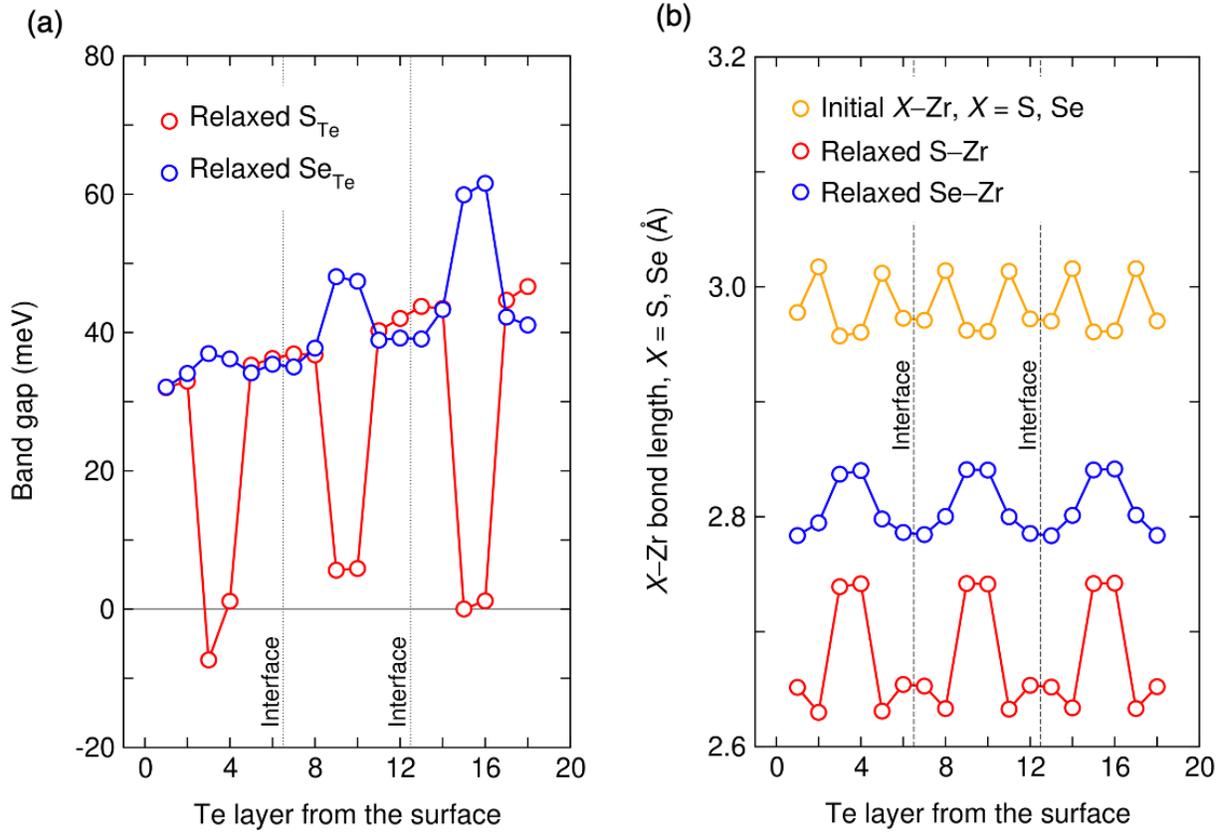

**Figure 2.** The Te-layer in ZrTe$_5$ for isovalent supfur doping (S$_{Te}$) and isovalent selenium doping (Se$_{Te}$) is compared in (a) with the bandgap and, in (b), with the bond length d$_{S-Te}$ and d$_{Se-Te}$. The trapezoidal pattern evident in both sets of trendlines can be related to the repetition of the CAP SSE at layers $(3 + 6n)$ or $(4 + 6n)$ $\forall\, n \in 0, 1,$ and $2$. The bandgap was derived by occupancy analysis at the $\bar{\Gamma}$, see Refs. [24,28].

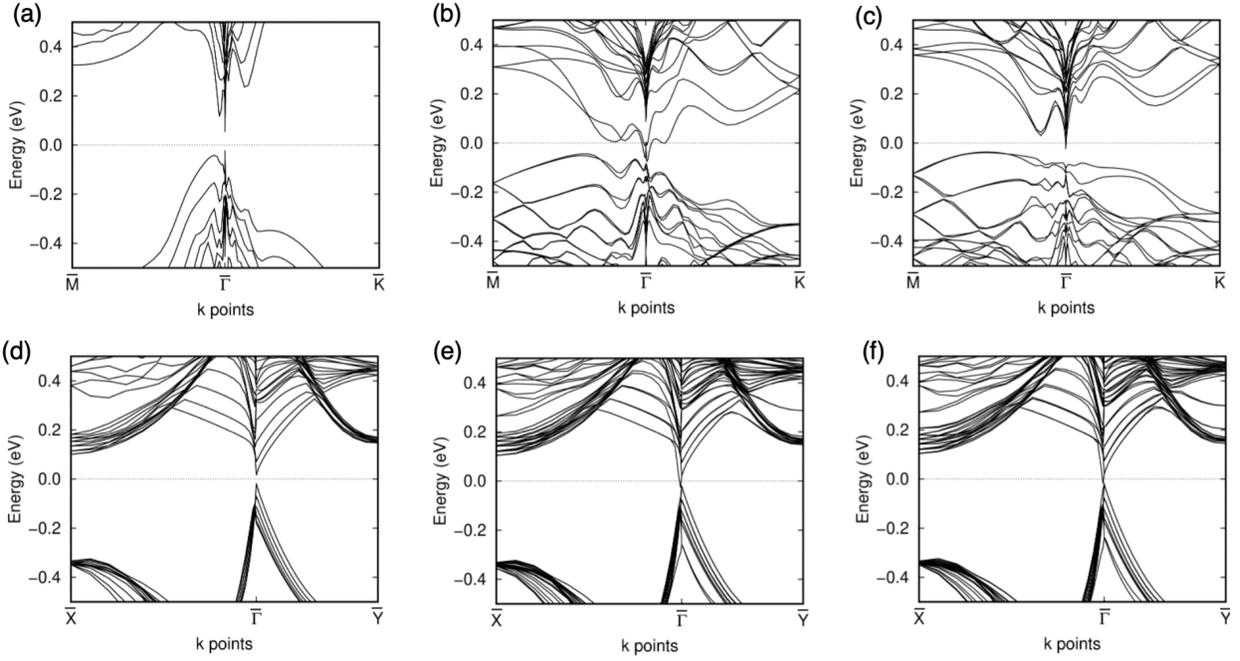

**Figure 3.** Band structures for $As_2Te_3$ in (a-c) and $ZrTe_5$ (d-f). The undoped and 100% sulfur-doped cases for Te-layer-site 5 and Te-layer-site 6 in $As_2Te_3$ are represented in (a), (b), and (c), respectively. The undoped and sulfur-doped cases on the Te-site of layer 3 and atomic layer 4 (the CAP layers) of $ZrTe_5$ are represented in (d), (e), and (f), respectively. In both cases, sulfur doping achieves total or significant bandgap closure (see Table 1 for cross-reference).